\documentclass[,final            
  ]
  {aipproc}

\layoutstyle{6x9}

\begin{document}

\title{Study of shape transitions in $N\sim90$ isotopes with beyond mean field calculations}


\author{Tom\'as R. Rodr\'iguez}{
  address={Departamento de F\'isica Te\'orica C-XI, Universidad Aut\'ononoma de Madrid, 28049, Madrid, Spain}
}

\author{J. L. Egido}{
  address={Departamento de F\'isica Te\'orica C-XI, Universidad Aut\'ononoma de Madrid, 28049, Madrid, Spain}
}

\begin{abstract}
We study the spherical to prolate-deformed shape transition in $^{144-158}$Sm and $^{146-160}$Gd isotopes with modern calculations beyond the mean field with the Gogny D1S force. We compare the results with the shape-phase transition predicted by the collective hamiltonian model and with the experimental data. Our calculations do not support the existence of a first order phase transition in these isotopic chains in the viewpoint of the Bohr hamiltonian neither the interpretation of the nuclei $N=90$ as critical points. 
\end{abstract}

\maketitle

Recently, the study of the shape transitions in isotopic chains and its connection with possible quantum phase transitions has been an intensive field of research.  The renewal interest about this topic is partly due to the development of analytic solutions for the critical points that correspond to phase transitions between spherical-prolate deformed (X(5)) and spherical-$\gamma$-soft (E(5)) systems in the context of the Bohr hamiltonian \cite{Iach1,Iach2}. In particular, transitions between nuclei with vibrational spectra to nuclei with a rotational character have been observed experimentally in the $N\sim90$ stable isotopes. Furthermore, the transitional nuclei  $^{150}$Nd,  $^{152}$Sm and $^{154}$Gd have been proposed as the empirical realization of the critical points of first-order phase transitions due to the good agreement between the experimental data and the predictions given by the X(5) model \cite{Kru1,Cast1,Ton1}. In this contribution we study the spherical-prolate deformed shape transition in the isotopic chains corresponding to $_{62}$Sm and $_{64}$Gd with state-of-the-art beyond mean field calculations using the Gogny D1S interaction. This study extends the analysis performed for the Nd isotopes in the Ref. \cite{Rod1} to the Sm and Gd nuclei. We will compare the theoretical results with the experimental data and also we will analyze the possible existence of phase transitions in this region of the nuclear chart.\\  
\begin{figure}
\centering
\includegraphics[width=\linewidth]{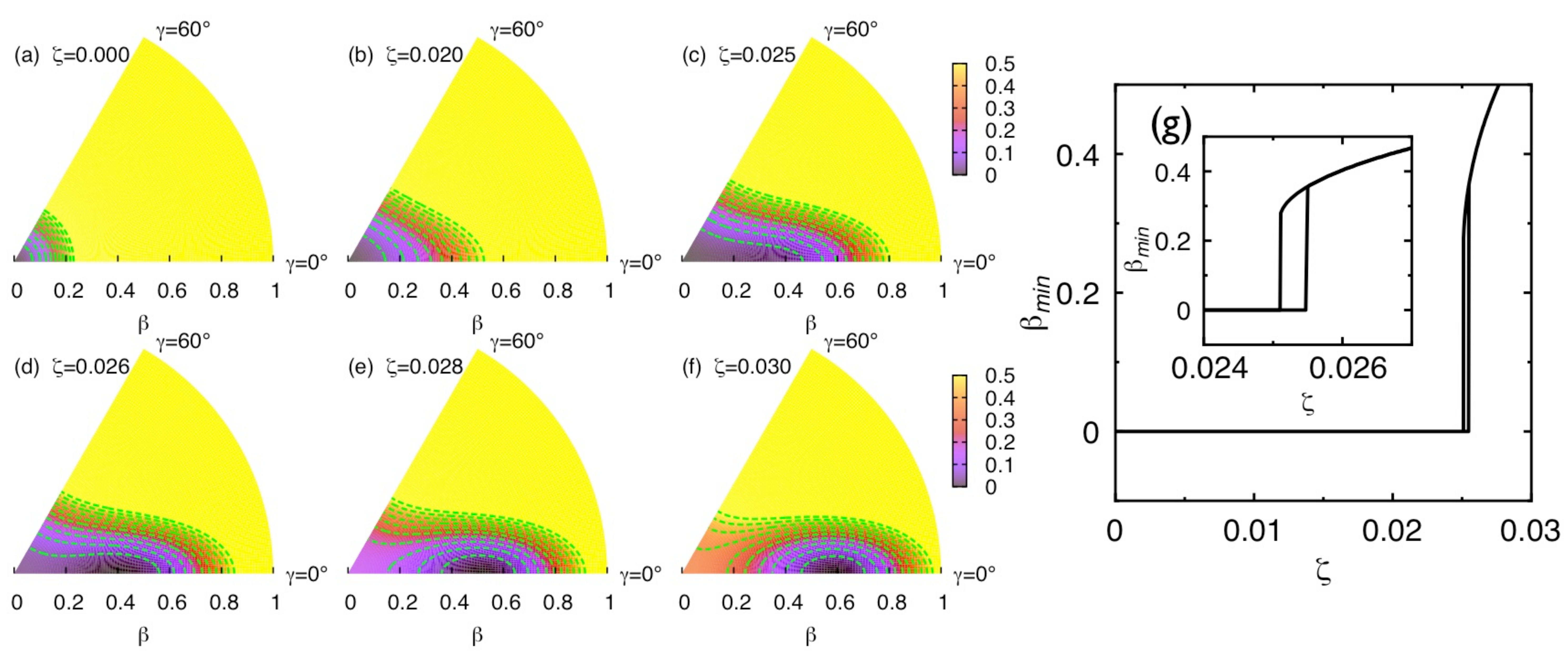} 
\caption{(a)-(f) Potential energy $V(\beta,\gamma)$ (Eq. \ref{bohr}) given by the collective model to describe the spherical to prolate deformed transition for different values of the control parameter $\zeta$. Contour lines are plotted each 0.05 arbitrary units, the energy is set to zero at the minimum of each surface. (g) Value of the $\beta$ deformation that corresponds to the minimum of (Eq. \ref{bohr}) as a function of $\zeta$. The inset shows the transitional region in detail.}
\label{Fig1}
\end{figure}
The description of these shape transitions as shape-\textit{phase} transitions has been performed in the context of the Bohr collective hamiltonian. In particular, the spherical-prolate deformed transition is obtained from the following potential in the $(\beta, \gamma)$ plane, deduced from the classical limit of the IBM model \cite{Iach2}:  
\begin{equation}
\label{bohr}
V(\beta,\gamma)=\frac{N\beta^{2}}{1+\beta^{2}}\left[1+\frac{5}{4}\zeta\right]-\frac{N(N-1)}{(1+\beta^{2})^{2}}\zeta\left[4\beta^{2}+2\sqrt{2}\beta^{3}\mathrm{cos} 3\gamma+\frac{1}{2}\beta^{4}\right]
\end{equation}
where $N$ is the number of valence pairs --we take $N=10$ as in Ref. \cite{Iach2}-- and $\zeta$ ($0\leq\zeta\leq1$) is the continuous control parameter that is used to simulate the shape transition. In Figs. \ref{Fig1}(a-f) the evolution of this potential in the $(\beta,\gamma)$ plane as a function of the parameter $\zeta$ is shown. For $\zeta=0$ a spherical minimum is obtained and the potential is very stiff in the $\beta$ direction. Increasing the value of $\zeta$ the potential softens in the trajectory $(\beta,\gamma=0^{\circ})$, emerging for $\zeta=0.025$ a deformed minimum that coexists with the spherical one in a small range of $\zeta$ values (Fig. \ref{Fig1}(g)). For higher values of the control parameter a single deformed minimum is obtained and the $\beta=0$ point turns into a saddle point. Precisely the value of $\zeta$ for which both minima are degenerated defines the critical point of the phase transition (see Fig. \ref{Fig4}(a)). \\
\begin{figure}
\centering
\includegraphics[width=\linewidth]{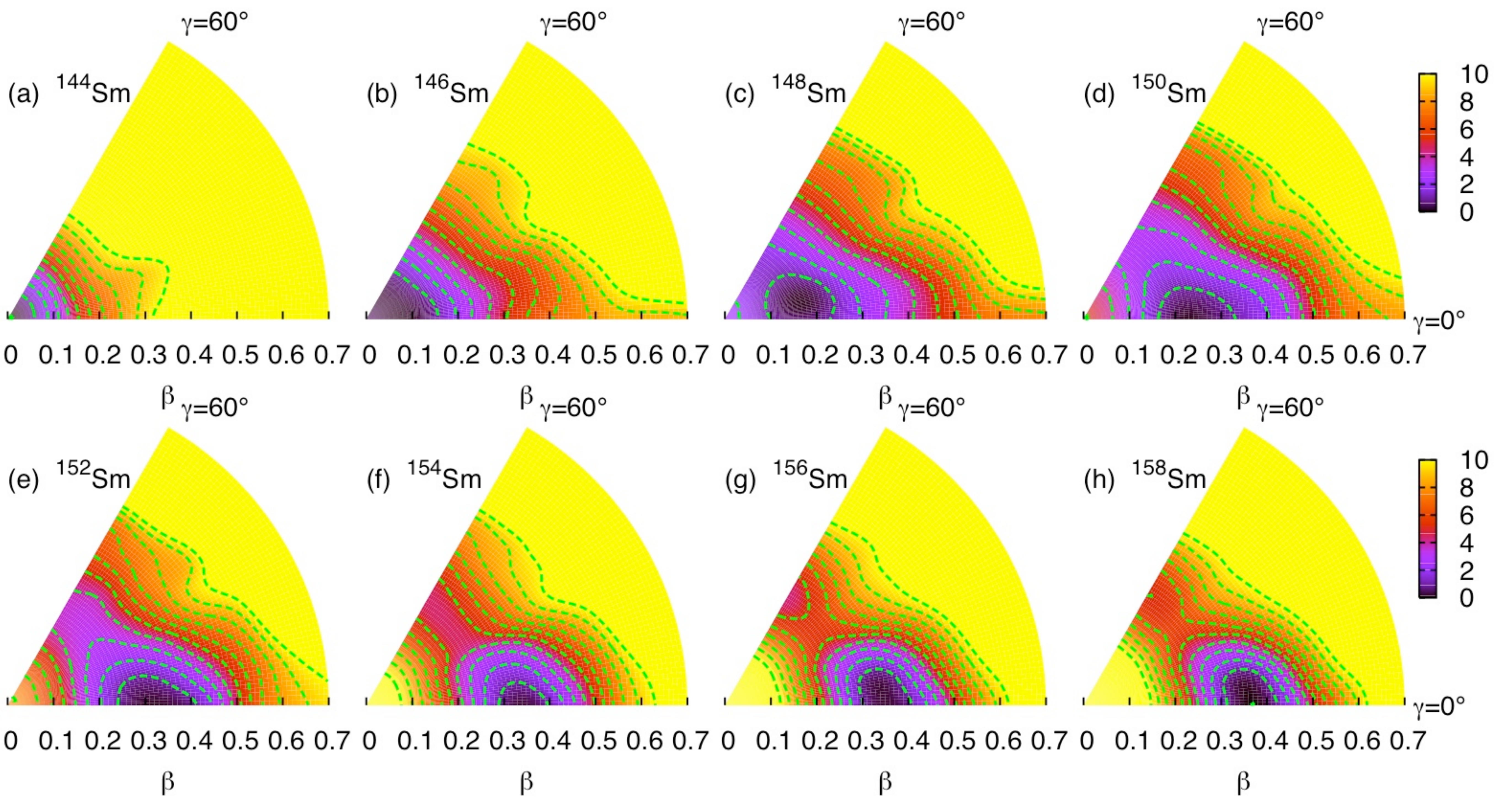} 
\caption{(a)-(h) Particle Number Projected Potential Energy Surfaces $E^{N,Z}(\beta,\gamma)$ (Eq. \ref{PES_VAP}) calculated with the Gogny D1S force for $^{144-158}_{62}$Sm isotopes. Contour lines represent a step of 1MeV and the energy is set to zero at the minimum of each surface.}
\label{Fig2}
\end{figure}
\begin{figure}
\centering
\includegraphics[width=\linewidth]{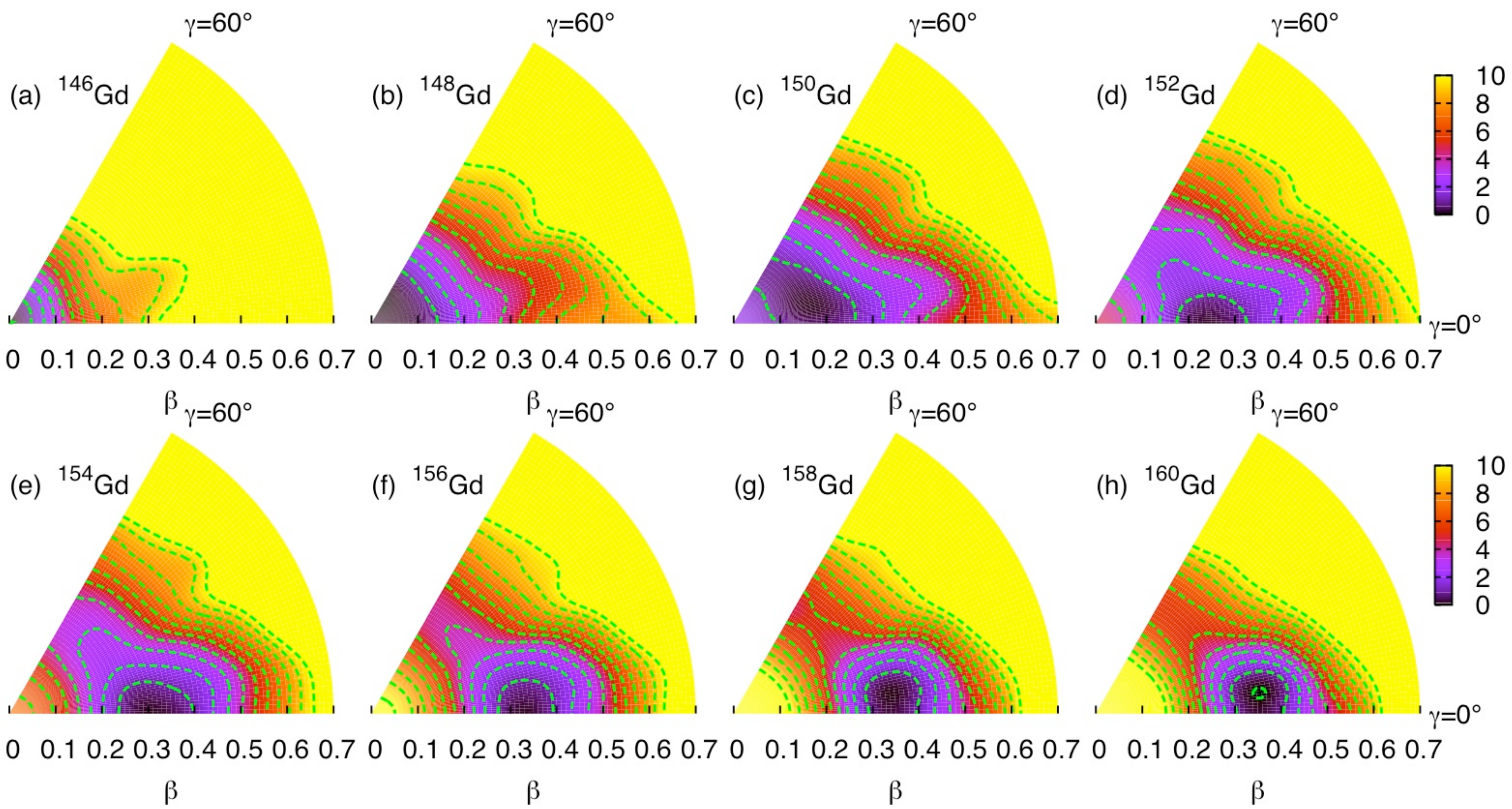} 
\caption{Same as Fig. \ref{Fig2} but for $^{146-160}_{64}$Gd isotopes.}
\label{Fig3}
\end{figure}
We now show the results of the beyond mean field calculations performed with the Gogny D1S force \cite{BERGNPA84}. The microscopic equivalent to the Eq. \ref{bohr} are the Potential Energy Surfaces (PES) \cite{Ring1} which are calculated for each nucleus evaluating the following expectation value:
\begin{equation}
E^{N,Z}(\beta,\gamma)=\frac{\langle\Phi^{N,Z}(\beta,\gamma)|\hat{H}|\Phi^{N,Z}(\beta,\gamma)\rangle}{\langle\Phi^{N,Z}(\beta,\gamma)|\Phi^{N,Z}(\beta,\gamma)\rangle}
\label{PES_VAP}
\end{equation}
where $|\Phi^{N,Z}(\beta,\gamma)\rangle$ are HFB-type wave functions that are projected -before the variation- onto good particle numbers ($N,Z$) and constrained to the deformation parameters ($\beta,\gamma$) -- see Refs. \cite{Rod1,Rod2} and references therein for more details.\\  
In Figs. \ref{Fig2}-\ref{Fig3} the PES with number of neutrons between $N=82-96$ are represented for Sm and Gd nuclei respectively. Similar results for Nd isotopes are obtained in Ref. \cite{Rod1} and are not shown here.
The three elements exhibit a similar general behavior. We observe that increasing the number of neutrons a shape transition from spherical shapes --obtained for the semi-magic $N=82$ nuclei-- to well prolate deformed nuclei is given. However, the manner in which these transitions are produced is different to the results of the collective hamiltonian shown in Fig. \ref{Fig1}. On the one hand, the microscopic calculations reveal the importance of the triaxial degree of freedom in the transitional nuclei $N=86,88,90$ (Figs. \ref{Fig2}-\ref{Fig3}(c-e)). Here the PES are soft in the $\gamma$ direction in contrast to the description given by the collective hamiltonian, where the degeneracy is produced along the $\gamma=0$ trajectory. On the other hand, none of the PES obtained with the Gogny force shows the structure of a critical point in the sense that we do not obtain any flat potential in the $(\beta, \gamma=0)$ direction with two degenerated coexisting minima, one of them spherical and the other one prolate deformed.
\begin{figure}
\centering
\includegraphics[scale=0.35]{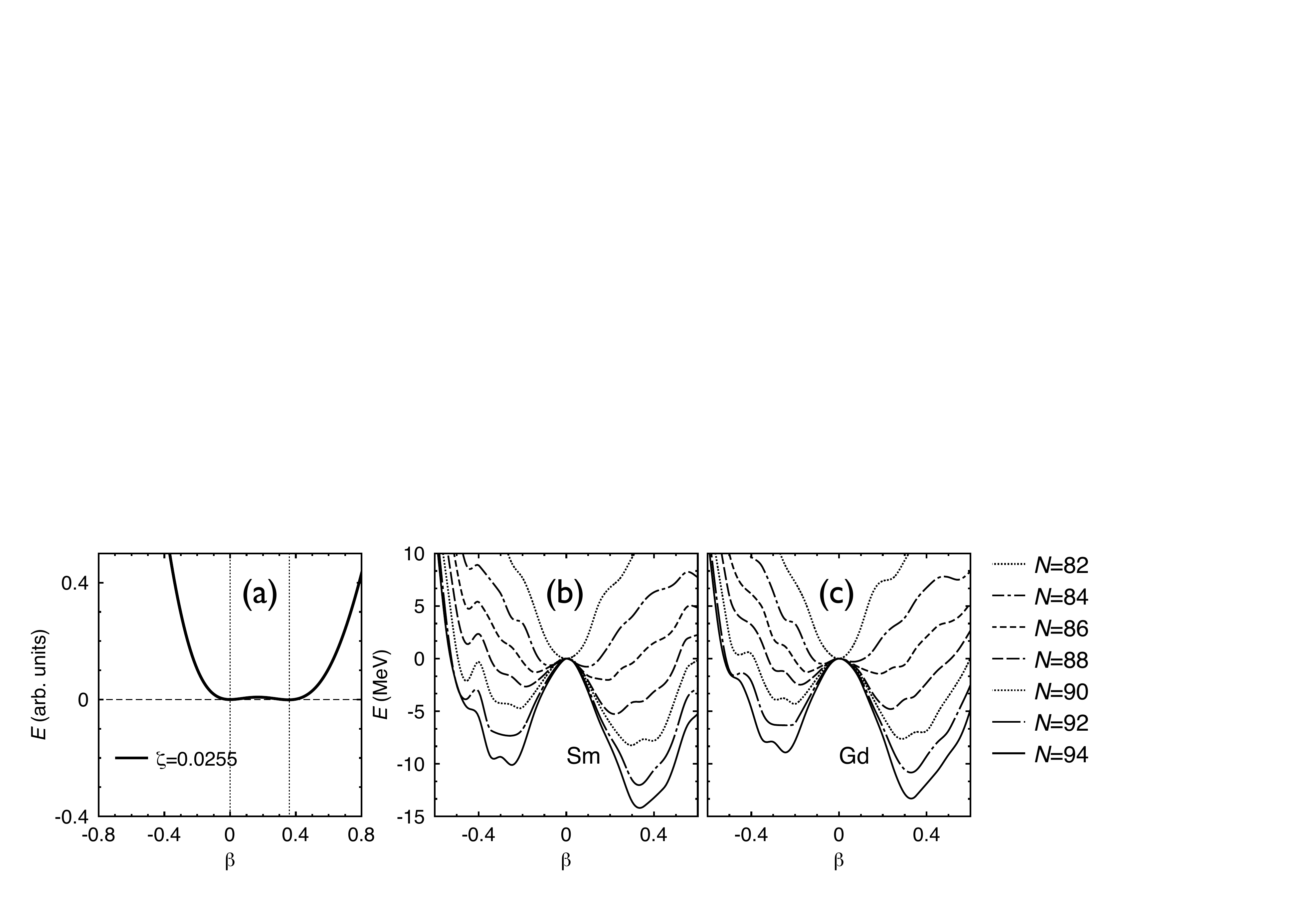} 
\caption{(a) Potential energy $V(\beta,\gamma=0,60)$ (Eq. \ref{bohr}) in the critical point as a function of $\beta$. Vertical lines mark the position of the two minima. (b)-(d) Particle number projected potential energy surfaces along the axial trajectory calculated with the Gogny D1S force for  $^{144-156}$Sm and $^{146-158}$Gd isotopes respectively. The zero of the energy is chosen at $\beta=0$.}
\label{Fig4}
\end{figure}
This characteristic can be clearly seen in Fig. \ref{Fig4}, where the PES along the axial direction are represented for Sm (Fig. \ref{Fig4}(b)) and Gd (Fig. \ref{Fig4}(c)) nuclei as well as the collective potential that corresponds to the critical point in Eq. \ref{bohr} (Fig. \ref{Fig4}(a)). We observe that spherical minima are obtained only for the shell closure $N=82$ nuclei. For the rest of isotopes we obtain oblate and prolate deformed minima where both the value of the $\beta$ for which those minima appear and also the height of the barrier between them are increasing with higher values of the number of neutrons. Obviously, most of the oblate minima are indeed saddle points in the $(\beta, \gamma)$ plane as it has been shown before. \\
\begin{figure}[b]
\centering
\includegraphics[scale=0.35]{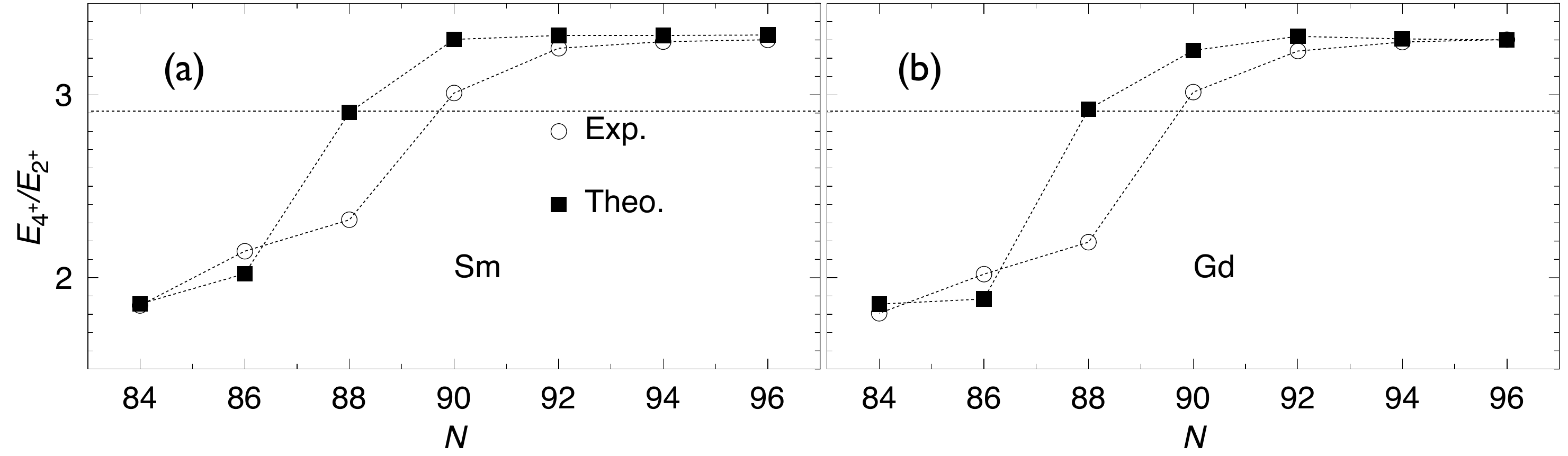} 
\caption{Ratio between the energy of the first 4$^{+}$ and $2^{+}$ excited states for (a) Sm and (b) Gd isotopes. The dash line is the X(5) prediction.}
\label{Fig5}
\end{figure}
In order to obtain theoretical predictions not only for the PES but also for the spectroscopy we carry out calculations within the framework of the Generator Coordinate Method (GCM) with the wave functions $|\Phi^{N,Z}(\beta,\gamma=0,60)\rangle$ also projected onto good angular momentum $J$. Hence, the resulting wave function can be expressed as:
$|\Psi^{N,Z,J,\sigma}\rangle=\int f^{N,Z,J,\sigma}(\beta)\hat{P}^{J}|\Phi^{N,Z}(\beta)\rangle d\beta$
where $\hat{P}^{J}$  is the angular momentum projection operator and the coefficients $f^{N,Z,J,\sigma}$ and the energies of the collective states $E^{N,Z,J,\sigma}$ are found by solving the Hill-Wheeler-Griffin equation \cite{Ring1} 
\begin{equation}
\int \left(\mathcal{H}^{N,Z,J}(\beta,\beta')-E^{N,Z,J,\sigma}\mathcal{N}^{N,Z,J}(\beta,\beta')\right)f^{N,Z,J,\sigma}(\beta')d\beta'
\end{equation}
being $\mathcal{H}^{N,Z,J}(\beta,\beta')$ and $\mathcal{N}^{N,Z,J}(\beta,\beta')$ the projected hamiltonian and norm overlaps respectively \cite{Ring1}.
The calculations are performed with intrinsic axially symmetric wave functions due to the huge computational cost of the full triaxial case. The first microscopic calculations performed to study the shape transitions of the $^{148-150-152}$Nd isotopes using a method very similar to the described above were carried out by Nik\u{s}i\'c and collaborators with a relativistic interaction \cite{Niksic1}.\\ 
The results obtained for the ground state band ($\sigma=1$) are compared with the experimental data in Fig. \ref{Fig5}. Here we represent the ratio $r_{4,2}=E(4^{+}_{1})/E(2^{+}_{1})$ for Sm and Gd isotopes. We observe both in the theoretical and the experimental results that increasing the number of neutrons in the system the ratio varies from $r_{4,2}\approx2$ --vibrational spectrum-- to values close to $r_{4,2}\approx3.3$ that correspond to a rotational spectrum. This behavior is a signature of the presence of a shape transition in this region. We obtain a good agreement between the data and the theoretical results in the vibrational and rotational limits while some deviations are observed in the transitional isotopes $N=86-90$. Precisely it is in these nuclei where the triaxial degree of freedom --neglected in this approximation-- is more relevant. Furthermore, it is known that the octupole deformation --which violates parity symmetry-- plays also an important role in these nuclei and could affect the quality of the description in this region \cite{Garr1}.\\
Finally, as it was mentioned above, the $^{150}$Nd, $^{152}$Sm and $^{154}$Nd nuclei show experimentally excitation spectra similar to the one predicted by the X(5) solution \cite{Kru1,Cast1,Ton1}. However, in the axial GCM calculations these isotopes possess a rotational character while the $N=88$ nuclei are closer to the X(5) prediction --see horizontal dotted line in Fig. \ref{Fig5}. This similarity is given not only in the $r_{4,2}$ ratio but also in the higher excited states and transition probabilities $B(E2)$ \cite{Rod1} (not shown). Nevertheless, these excitation spectra, though reproducing the X(5) spectrum, are extracted from intrinsic potentials that have not a clear connection to the X(5) potential (Fig. \ref{Fig5}). Also Nik\u{s}i\'c and collaborators \cite{Niksic1} obtain an excellent agreement with the experimental data for $^{150}$Nd with an intrinsic potential different to the one of the critical point.\\
In summary, we have studied the spherical-prolate deformed shape transition in Nd, Sm and Gd isotopes with beyond mean field calculations using the Gogny D1S force. We have compared the results with the description of a first order shape-phase transition given by the collective hamiltonian showing that, contrary to this model, the triaxial degree of freedom plays an important role. Also the critical point potential is not observed. Both the vibrational and rotational empirical limits are obtained with GCM calculations although the agreement with the experimental data is worse in the transitional region, where some relevant degrees of freedom are neglected. Work is in progress to include them in future analyses.  

\begin{theacknowledgments}
The authors acknowledge financial support from the Spanish Ministerio de Educaci\'on y Ciencia
under contract FPA2007-66069 and by the Spanish Consolider-Ingenio 2010 
\end{theacknowledgments}

\bibliographystyle{aipproc}   


\end{document}